\documentclass[twoside]{article}

\usepackage{graphicx}
\usepackage{a4wide}
\usepackage{mathptm}

\title{Power spectrum and 
Intermittency of the Transmitted Flux of QSOs' Ly$\alpha$ Absorption Spectra}

\author{Priya Jamkhedkar\footnote{Department of
       Physics, University of Arizona, Tucson, AZ 85721},
       Long-Long Feng\footnote{Center for Astrophysics,
       University of Science and Technology of China, Hefei,
       Anhui 230026, P.~R.~China, and National Astronomical
       Observatories Chinese Academy of Science, Chao-Yang District,
       Beijing 100012, P.~R.~China},
       Wei Zheng\footnote{Department of Physics and Astronomy,
       The Johns Hopkins University, Baltimore, MD 21218},\\
        David Kirkman\footnote{Center of Astrophysics and Astronomy,
        University of California, San Diego, CA 92093},
        David Tytler\footnote{Center of Astrophysics and Astronomy,
        University of California, San Diego, CA 92093},
       Li-Zhi Fang\footnote{Department
        of Physics, University of Arizona, Tucson, AZ 85721}}%

\date{}

\begin{document}

\maketitle

\begin{abstract}

Using a set of 28 high resolution, high signal to noise ratio (S/N) 
QSO Ly$\alpha$ absorption spectra, we investigate the non-Gaussian features 
of the transmitted flux fluctuations, and their effect upon the 
power spectrum of this field. We find that the spatial distribution of the 
local power of the transmitted flux on scales $k \geq 0.05$~s/km is highly 
spiky or intermittent. The 
probability distribution functions (PDFs) of the local power are long-tailed. 
The power on small scales is dominated by small probability events,
and consequently, the uncertainty in the power spectrum of the transmitted 
flux field is generally large. This uncertainty arises due to the 
slow convergence of an intermittent field to a Gaussian limit required 
by the central limit theorem (CLT). To reduce this uncertainty, it 
is common to estimate the error of the power spectrum by selecting 
subsamples with an ``optimal" size. We show that this conventional 
method actually does not calculate the variance of the original 
intermittent field but of a Gaussian field. Based on the analysis of 
intermittency, we 
propose an algorithm to calculate the error. It is based on a bootstrap 
re-sampling 
among all independent local power modes. This estimation doesn't require 
any extra parameter like the size of the subsamples, and is sensitive to the 
intermittency of the fields. This method effectively reduces the uncertainty 
in the power spectrum when the number of independent modes matches the 
condition of the CLT convergence.

\end{abstract}


\section{Introduction}

Ly$\alpha$ absorption on the short wavelength side of Ly$\alpha$ 
emission in QSO spectra with high redshift indicates the presence of 
neutral hydrogen (HI) along the line of sight. The low column density 
($10^{13}$ to $10^{17}$ cm$^{-2}$) HI absorption is believed to 
be due to diffusely distributed intergalactic medium (IGM). The IGM 
mass field is passive with respect to the dark matter present in the sense 
that its gravitational clustering is dominated by the gravity of dark 
matter. The IGM is assumed
to be a good tracer of the underlying mass 
density distribution on scales larger than the thermal diffusion or Jeans 
length in the linear or even the nonlinear regime
 (Bi, 1993; Fang et al. 1993; Hui, \& Gnedin  1997; Nusser, \& Haehnelt 
1999). The statistical features of the QSO Ly$\alpha$ transmitted flux 
can be used to estimate the corresponding features of the underlying 
mass field. Consequently, the power spectrum of the transmitted flux 
fluctuations of high redshift QSOs' Ly$\alpha$ absorption spectrum has been 
extensively used for reconstructing the initially linear perturbations of 
the cosmic mass and velocity fields and discriminating among models of 
structure formation (Bi, Ge, \& Fang 1995; Bi, \& Davidsen 1997; Croft et al. 
1999, 2002; Feng, \& Fang 2000; McDonald et al 2000; Pando, Feng 
\& Fang 2001; Zhan, \& Fang 2002). 

Therefore, it is necessary to study the effects of the non-Gaussian features 
of the flux field upon the detection of the power spectrum of this field, 
and its error estimation. In this paper, we will focus on the effects 
of intermittency of the QSO Ly$\alpha$ transmitted flux on the
power spectrum estimation.  

The first indication of intermittency arises from the success of 
the lognormal model in explaining the 
QSO Ly$\alpha$ forests (Bi 1993; Bi, \& Davidsen 1997) as a lognormal field 
typically is intermittent. To approximately describe the evolution 
of the IGM by a stochastic Burgers' equation, the IGM mass and velocity fields 
are found to be lognormal or intermittent (Jones, 1999; Matarrese, \& Mohayaee 
2002). Intermittency has been detected with high resolution, high signal 
to noise ratio (S/N) samples of QSO Ly$\alpha$ absorption spectra 
(Jamkhedkar, 
Zhan, \& Fang 2000; Feng, Pando, \& Fang 2001; Zhan, Jamkhedkar, \& Fang 2001; Jamkhedkar 2002). 
Recently, the intermittent features have been further examined by measuring 
the structure function and intermittent exponent of 28 Keck HIRES QSO spectra 
(Pando et al. 2002). This work shows that the intermittent behaviour is 
significant on scales $k > 0.05$ s/km. 

A basic characteristic of an intermittent field is that the power of the
fluctuations of the flux field is concentrated in rare spikes which are 
randomly and widely scattered in space, with low power between 
the spikes. In this case, the power spectrum of an intermittent field is 
dominated by rare and improbable events (spikes). In other words, only
a small fraction of the total modes contribute to the power while 
most are inactive. Although an intermittent field is statistically 
homogeneous, the rare events lead to significant differences among samples 
from different parts of the universe when the spatial size of the region 
is not large enough to contain numerous spikes. Mathematically, this makes 
the central limit theorem less effective.  All these imply that the 
uncertainty in the power spectrum is directly related to the intermittency of
the field. 
   
Using the same samples studied by Pando et al. (2002), we 
will investigate the uncertainty of the power spectrum of QSO Ly$\alpha$ 
transmitted flux fluctuations.  
The paper is organised as follows. \S 2 briefly introduces the basic 
properties of an intermittent field. \S 3 presents our algorithm for 
calculating the power spectra with the discrete wavelet transform (DWT).
\S 4 describes the samples used for analysis. \S 5 discusses the basic 
properties of the DWT power spectrum of the samples. \S 6 addresses
the effect of the intermittent behaviour upon the power spectrum.
Finally, \S 7 summarises the results and conclusions.

\section{Intermittent field}

Most of this section has been presented in our previous work
(Pando et al 2002). For the paper to be self-contained, we repeat 
these results very briefly. The principal characteristic of an 
intermittent random field, like Ly$\alpha$ transmitted flux 
$F({ \lambda})$, is measured by the asymptotic behaviour of the ratio 
between the high- and low-order moments of the field defined as
\begin{equation}
\frac{\langle [F( x+ r)- F({ x})]^{2n}\rangle}
{[\langle [F({ x + r})- F({ x})]^2\rangle]^n}
\asymp\left \langle \frac{r}{L} \right \rangle^{\zeta},
\end{equation}
where $\langle...  \rangle$ is the average over the ensemble of fields,
and $L$ is the size of the sample. Eq.~(1) can be rewritten as
\begin{equation}
\frac{S^{2n}_r}{[S^{2}_r]^n} \propto
   \left(\frac{r}{L} \right )^{\zeta},
\end{equation}
where $S^{2n}_r$ is the structure function defined  by
\begin{equation}
S^{2n}_r = \langle|\Delta_r(x)|^{2n} \rangle.
\end{equation}
Here $\Delta_r(x)\equiv F(x+r)-F(x)$. 
$S^{2n}_r$ is the $2n^{\rm th}$ moment of the fluctuations of 
the field. Therefore, the ratio in eq.~(2) is the $2n^{\rm th}$ moment 
$S^{2n}$ normalised by the $n^{\rm th}$ power of the $2^{\rm nd}$ moment, 
or power, $S^2$.

A field is intermittent if the exponent $\zeta$ is negative on small 
scales $r$ (G\"artner, \& Molchanov, 1990; Zel'dovich, Ruzmaikin, \& 
Sokoloff, 1990). Intermittency is measured by the $n$- and $r$-dependencies 
of $\zeta$. For a Gaussian field,
\begin{equation}
\frac{S^{2n}_r}{[S^{2}_r]^n} = (2n-1)!!.
\end{equation}
This ratio is independent of the scale $r$, and therefore, the
intermittent exponent $\zeta=0$. Thus, a Gaussian field is not 
intermittent. Moreover, not all non-Gaussian fields are intermittent.  
Only fields, for which the ratio in eq.~(1) diverges as 
$r \rightarrow 0$, are intermittent.

For a lognormal field, the probability distribution function (PDF) 
of $\Delta^2_r(x)$ is 
\begin{equation}
P[\Delta^2_r(x)] =\frac{1}{2^{1/2}\pi^{1/2}\Delta^2_r(x)\sigma(r)}
 \exp\left \{ -\frac{1}{2}
\left (\frac {\ln\Delta^2_r(x)-\ln \Delta^2_r(x)_m}
{\sigma(r)}\right)^2  \right \},
\end{equation}
where $\Delta^2_r(x)_m$ is the {\it median} of $\Delta^2_r(x)$ 
(Vanmarcke 1983). The variance $\sigma(r)$ of $\ln \Delta^2_r(x)$
 may be a function of the scale $r$. Using eq.~(5), we have
\begin{equation}
\frac{S^{2n}_r}{[S^{2}_r]^n} = e^{(n^2-n)\sigma^2(r)/2}.
\end{equation}
The intermittent exponent of a lognormal field is then
\begin{equation}
\zeta \simeq (n^2-n)\sigma^2(r)/2\ln(r/L).
\end{equation}
Because $r < L$, $\zeta$ is negative. Therefore, a lognormal 
field is intermittent.

In comparison with the Gaussian result (eq.~(4)), the ratio of eq.~(6)
increases faster with $n$. This is due to the fact that a  
lognormal PDF is long-tailed. long-tailed PDF is a common property of 
intermittent fields. The property of $S^{2n}_r \gg [{S^{2}_r}]^n$ on small 
scales indicates that the field  contains  ``abnormal'' events of 
large density fluctuations $|\Delta_r (x)|$. The events of
big  $|\Delta_r(x)|$ correspond to a sharp increase or decrease of the field,
and constitute the long-tail of the PDF of $|\Delta_r(x)|$. 

\section{DWT algorithm of power spectrum of Ly$\alpha$ transmitted flux}

\subsection{The field of Ly$\alpha$ transmitted flux fluctuations}

The transmitted flux of a QSO absorption spectrum is given by 
$F(\lambda)= F_c(\lambda)e^{-\tau(\lambda)}$, where $F_c(\lambda)$ is the 
continuum, $e^{-\tau(\lambda)}$ the transmission, and $\tau(\lambda)$ the 
opacity. Since the observed data (\S 4) is already reduced by a continuum 
fitting, we have 
\begin{equation}
F(\lambda)=e^{-\tau(\lambda)} + n(\lambda),
\end{equation}
where the term $n(\lambda)$ describes the stochastic noise including
the Poisson noise of photon count. It satisfies the 
statistical properties 
\begin{equation}
\langle n(\lambda)\rangle =0, \hspace{5mm}
\langle n(\lambda)n(\lambda') \rangle
= \sigma^2(\lambda)\delta^K_{\lambda,\lambda'},
\end{equation}
where $\delta^K$ is the Kronecker Delta function.  
The ensemble average is over the density
fluctuations. Therefore, we have
$\langle{F(\lambda)}\rangle=\langle e^{-\tau(\lambda)}\rangle$.  
Generally speaking, $\langle{F(\lambda)}\rangle$ is 
still position-dependent, as it depends on the photoionization rate 
of HI, and the temperature of the IGM. Moreover, for a given 
scale $r$, all fluctuations on scales larger than $r$ around the position 
$\lambda$ act as a background for this fluctuation. This leads 
to a position dependent background. If one can assume that 
$\langle{F(\lambda)}\rangle$ is a constant, then we have 
$\langle{F(\lambda)}\rangle=
\overline{\langle e^{-\tau(\lambda)}\rangle }$, which is the mean 
transmission. 

The field of Ly$\alpha$ transmitted flux fluctuations are defined by
\begin{equation}
\delta^n(\lambda)=\frac{F(\lambda)-\langle F(\lambda)\rangle }
                    {\langle F(\lambda)\rangle },
\end{equation}
or
\begin{equation}
\delta(\lambda)=F(\lambda)-\langle F(\lambda)\rangle.
\end{equation}
$\delta^n(\lambda)$ and $\delta(\lambda)$ are different by a 
normalisation factor $\langle{F(\lambda)}\rangle$. If 
$\langle{F(\lambda)}\rangle$ is $\lambda$-independent, the unnormalised and 
normalised fields $\delta^n(\lambda)$ and $\delta(\lambda)$  
differ only by a constant. If $\langle{F(\lambda)}\rangle^2$ doesn't 
contain small scale fluctuations, then one still can treat 
$\langle{F(\lambda)}\rangle^2$ as a constant when studying the fields 
on small scales. 

\subsection{DWT variables of Ly$\alpha$ transmitted flux field}

In order to easily illustrate the effect of intermittency on the 
power spectrum, we calculate the power spectrum of Ly$\alpha$ transmitted 
flux with the 
discrete wavelet transform (DWT). 

We use $x_1$ and $x_2$ to denote the spatial range of a flux field 
corresponding to the 
wavelength range from $\lambda_1$ and $\lambda_2$. To implement a DWT 
scale-space decomposition of a flux field $F(x)$, we first chop the 
spatial range $L=x_2-x_1$ of the 1-D sample into $2^j$ subintervals 
labelled with $l=0, ..., 2^j-1$. Each subinterval spans a spatial range 
$L/2^j$. The subinterval $l$ is from $x_1+ Ll/2^j$ to $x_1 + L(l+1)/2^j$. 
The index $j$ can be an integer. Thus, we decompose the space $L$ 
into cells $(j,l)$, where $j$ denotes the scale $L/2^j$, and $l$ 
the spatial range $[x_1+Ll/2^j, x_1+L(l+1)/2^j]$. 

Corresponding to each cell (or mode) $(j,l)$ , there is a scaling function 
$\phi_{jl}(x)$, and a wavelet function $\psi_{jl}(x)$, which are the  
orthogonal and complete basis for the scale-space decomposition.
The most important property of the DWT basis is its locality in both 
scale and physical (or redshift) spaces. The scaling function 
$\phi_{jl}(x)$ is a window function on scale $j$ and at the position 
$l$. The wavelet function $\psi_{jl}(x)$ is admissible (Daubechies 1992), 
i.e., $\int \psi_{jl}(x)dx=0$. Therefore, it measures the 
fluctuation on scale $j$ and at position $l$. All wavelets with compactly 
supported bases will produce similar results. We will use Daubechies 
4 (D4) basis below, for its ease in numerical calculations among the 
compactly supported orthogonal bases. The D4 scaling function 
$\phi_{jl}(x)$, wavelet $\psi_{jl}(x)$ and their Fourier 
transforms $\hat{\phi}_{jl}(n)$, $\hat{\psi}_{jl}(n)$ are shown in 
Yang et al (2001b). 

Subjecting a transmitted flux $F(x)$ to the DWT, we have (Daubechies 
1992; Fang, \& Thews 1998)
\begin{equation}
F(x) = \sum_{l=0}^{2^j-1}\epsilon^F_{jl}\phi_{jl}(x)+
 \sum_{j'=j}^{J-1} \sum_{l=0}^{2^{j'}-1}
  \tilde{\epsilon}^F_{j'l} \psi_{j'l}(x),
\end{equation}
where $J$ is given by the finest scale (pixel resolution) of the sample, 
i.e., $\Delta x=L/2^J$ and $j$ is the scale we want to study. The field 
now is described by DWT variables $\epsilon^F_{jl}$ and 
$\tilde{\epsilon}^F_{jl}$, which are called scaling function coefficient 
(SFC) and wavelet function coefficient (WFC), respectively.

The SFC is given by projecting $F(x)$ onto $\phi_{jl}(x)$
\begin{equation}
\epsilon^F_{jl}=\int F(x)\phi_{jl}(x)\:dx.
\end{equation}
The SFC $\epsilon^F_{jl}$ is proportional to the mean field at the 
mode $(j,l)$. The WFC is obtained by projecting $F(x)$ onto 
$\psi_{jl}(x)$
\begin{equation}
\tilde{\epsilon}^F_{jl}= \int F(x) \psi_{jl}(x)\:dx.
\end{equation}
The WFC $\tilde{\epsilon}^F_{j l}$ is basically the difference between 
$F(x)$ and $F(x+r)$, where $x \simeq x_1+ lL/2^j$ and $r\simeq L/2^j$.
Thus, the WFC can be used to replace the variable $F(x+r)-F(x)$.
Eq.~(3) can then be rewritten as 
\begin{equation}
S^{2n}_j = \langle|\tilde{\epsilon}^F_{jl}|^{2n} \rangle.
\end{equation}
If the ``fair sample hypothesis'' (Peebles 1980) holds, one
can replace the ensemble average in eq.~(15) by a spatial average. We then have
\begin{equation}
S^{2n}_j = \frac{1}{2^j}\sum_{l=0}^{2^j-1}|\tilde{\epsilon}^F_{jl}|^{2n}.
\end{equation}

\subsection{Power spectra of Ly$\alpha$ transmitted flux}

For a transmitted flux fluctuation field $\delta^n(x)$, the two point 
correlation function is defined by 
\begin{equation}
\xi^n(\Delta x)=\langle \delta^n(x_1)\delta^n(x_2)\rangle,
\end{equation}
where $\Delta x= x_2-x_1$. Similarly, one can define a two point correlation 
function of the unnormalised field $\delta(x)$ as
\begin{equation}
\xi(\Delta x) = \langle \delta( x_1) \delta( x_2) \rangle.
\end{equation}
The Fourier counterparts of $\xi^n(\Delta x)$ and $\xi(\Delta x)$ are, 
respectively, the normalised and unnormalised power spectra, $P^n(k)$ 
and $P(k)$. 

The algorithm of the power spectrum for both $P^n(k)$ or $P(k)$ in the 
DWT representation has been developed (Pando \& Fang 1998; Fang, \& Feng 
2000; Yang et al. 2001a, 2001b and Jamkhedkar, Bi, \& Fang 2001).
We will study only the unnormalised power spectrum $P(k)$ below. If 
$\langle{F(\lambda)}\rangle$ is constant, the 
normalised and unnormalised power spectra $P^n(k)$ and $P(k)$ 
differ only by a constant factor $\langle{F(\lambda)}\rangle^2$. In this case,
all the results for $P(k)$ also hold for $P^n(k)$. 
Problems with $P^n(k)$ caused by a non-constant 
$\langle{F(\lambda)}\rangle$ will be discussed in the conclusions.

The power spectrum in the DWT variables is given by  
\begin{equation}
P_j=\frac{1}{2^j}\sum_{l=0}^{2^j-1}(\tilde{\epsilon}^F_{jl})^2
   - \frac{1}{2^j}\sum_{l=0}^{2^j-1}(\tilde{\epsilon}^n_{jl})^2.
\end{equation}
The first term on the r.h.s. of eq.~(19) is the power 
of $\delta(x)$ on scale $j$. The second term on the r.h.s. of 
eq.~(19) is due to the noise given by
\begin{equation}
(\tilde{\epsilon}^n_{jl})^2 = \int\sigma^2(x)\psi_{jl}^2(x)\:dx.
\end{equation}
Comparing eq.~(19) with eq.~(16), we have $P_j = S_j^2$. Therefore,
the power spectrum actually is the second moment of the PDF of
$\tilde{\epsilon}^F_{jl}$.
 
It has been shown in general that the DWT power spectrum $P_j$ of a 
random field $\delta(x)$ is related to the Fourier power spectrum $P(n)$ 
of the field as
\begin{equation}
P_j = \frac{1}{2^j} \sum_{n = - \infty}^{\infty}
 |\hat{\psi}(n/2^j)|^2 P(n),
\end{equation}
where $n$ is related to the wavenumber by $k = 2 \pi n/L$. Therefore, the 
DWT power $P_j$ is a banded Fourier power spectrum. Since for D4 
wavelet $|\hat{\psi}(\eta)|^2$ has peak at $\eta \simeq 1$, the band 
$j$ is around the wavenumber $k = 2 \pi n/L \simeq 2\pi 2^j/L$. 

The difference between the DWT mode $(j,l)$ and the Fourier mode $k$ 
should be emphasised. For a given $j$, the size of a cell (mode) is
$L/2^j$, which corresponds to wavenumber $k=2\pi 2^j/L$. However, the 
wavelet function $\psi_{jl}(x)$ is localised within $L/2^j$, and, therefore,
the uncertainty relation $\Delta x \Delta k =2\pi$ gives 
$\Delta k =2\pi /(L/2^j)=k$. That is, a DWT mode with $j$ actually
corresponds to a band of Fourier modes $k \pm k/2$, i.e., the relevant band
is $\Delta k/k =\Delta \ln k \simeq 1$. This banding is optimised in the 
sense that to detect small scale fluctuations 
(larger wavenumber $k$), the size of the pieces $\Delta x$ is chosen 
to be smaller, while to detect large scale fluctuations (smaller wavenumber), 
the size of the pieces $\Delta x$ is chosen to be larger.
We sometimes characterise a DWT scale $j$ by the Fourier mode $k$. In this 
case, the $k$ is given by the maximum $\eta_{max}$ of the Fourier transform 
$|\hat\psi_{j l}(\eta)|$.

\section{Samples}
 
\subsection{Observed data}

The observational data used in our study consists of 28 Keck HIRES QSO 
spectra (Kirkman, \& Tytler 1997). The QSO emission redshifts 
cover a redshift range from 2.19 to 4.11.  For each of the 28 QSOs, the 
data are given in form of pixels ($i=1,2,\ldots$) with wavelength 
$\lambda_i$, flux $F(\lambda_i)$ and noise $\sigma(\lambda_i)$. The 
noise accounts for the Poisson fluctuations in the photon count, 
the noise due to the 
background and the instrumentation. The continuum of each spectrum 
is given by IRAF CONTINUUM fitting. 

For our purpose, the useful wavelength region is from the Ly$\beta$ 
emission to the Ly$\alpha$ emission, excluding a region of about 0.06 
in redshift close to the quasar to avoid any proximity effects. In this 
wavelength range, the number of pixels of the data is about 
$1.2\times 10^4$ for 
each spectrum. Fig.~1 shows the wavelength range for 30 QSOs. Since the 
data of Q0241-0146 has about 44\% and, Q1330+0108 has about 
27\% pixels with S/N $<$ 3, we will not 
use these QSOs in our analysis. We use only the 28 QSO 
forest samples. 

For each bin in this data set, the ratio $\Delta \lambda/\lambda$ is 
constant, i.e., $\Delta \lambda/\lambda \simeq 13.8\times 10^{-6}$, or 
$\delta v \simeq 4.01$~km/s, and, therefore, the resolution is 
about 8 km/s. The distance between $N$ pixels in the units 
of the local velocity scale is given by 
$\Delta v=2c(1-\exp{[-(1/2)N\delta v/c]})$~km/s, or wavenumber
$k=2\pi /\Delta v$ s/km. We use only $2^{13}=8192$ pixels of each 
spectrum. Thus, each cell on the DWT scale $j$ corresponds to 
$N=2^{13-j}$ pixels. 

This data set has been used for the Fourier power spectrum analysis 
(Croft et al 1999, 2002). They concluded that on scales 
$k \leq 0.15$~s/km the contamination of noise is small. In order to 
easily compare our results with the Fourier analysis, we concentrate
mainly on the DWT scales $j \leq 10$, or bin number
$N \geq 8$, $k \leq 0.2$~s/km, or $\Delta v \geq 32~{\rm km/s}$.

The samples are contaminated by metal lines. It is generally believed 
that metal lines are narrow with a Doppler parameter 
$b \leq 15$~km/s (Rauch, et al 1997; Hu et al 1995). We identified the
big spikes on the scale $\sim 32$~s/km for the ten QSOs 
(Q0014+8118, Q0054-2824, Q0636+6801, Q0642+4454, Q0940-1050, Q1017+1055, 
Q1103+6416, Q1422+2309, Q1425+6039, Q1759+7529) and checked if these are 
related to metal lines. To estimate the 
effect of metal lines, we compare the statistical results of 
metal-line-removed samples with those without removing metal lines.

In our analysis, we sometimes use the 28 QSO transmissions 
individually, i.e.\ calculate the statistics of the transmission over 
each QSO separately, and sometimes all the transmissions are treated  
together. In the latter case, we divide the data into 12 redshift ranges 
from $z=1.6 + n\times 0.20$ to 1.6 + $(n+1)\times$ 0.20 where $n=0,\ldots,11$. 
All the transmission flux in a given redshift range forms an ensemble. 
Note that the number of data points in each redshift range is different. 

The mean flux $\langle F \rangle$ of these samples at $z=1.6$ is 
$\sim 0.85$, and decreases to $\sim 0.5$ at $z=3.5$. In each redshift 
range, the dispersions of $\langle F \rangle$ are about 10-15\%.  

\subsection{Treatment of unwanted data}

Before calculating the DWT power spectrum of observed data, we discuss 
our method of treating unwanted data, including the pixels without data,
contamination of metal lines, etc. On an average the S/N ratio of the 
Keck spectra is high. The mean 1$\sigma$ uncertainty in the flux values 
$F$ relative to the continuum $F_c$ in the Ly$\alpha$ forest region is 
4\% on an average (Croft et al 2002). However, for some pixels, the $S/N$ 
is as low as about 1, such as pixels with negative flux. Most of these 
regions are saturated absorption regions. Although the percentage of pixels
within these regions is not large, they may introduce large uncertainties 
in the analysis. We must reduce the uncertainty given by low S/N pixels.  

The conventional technique of reducing these uncertainties is eliminating 
the unwanted bins and smoothly rejoining the rest of the forest spectra. 
However, by taking advantage of the localisation of wavelets we can use 
the algorithm of DWT denoising by thresholding (Donoho 1995) as follows 
\begin{enumerate}
\item Calculate the SFCs of both transmission $F(x)$ and 
noise $\sigma(x)$, i.e.
\begin{equation}
\epsilon^F_{jl}=\int F(x)\phi_{jl}(x)\:dx, \hspace{3mm}
\epsilon^N_{jl}=\int \sigma(x)\phi_{jl}(x)\:dx.
\end{equation}
\item Identify an unwanted mode $(j,l)$ using the threshold condition 
\begin{equation}
\left| \frac{\epsilon^F_{jl}} {\epsilon^N_{jl}} \right| <  f,
\end{equation}
where $f$ is a constant. This condition flags all modes with 
S/N less than $f$. We can also flag modes dominated by metal lines. 
\item Since all the statistical quantities in the DWT representation 
are based on an average over the modes $(j,l)$, we will skip all the 
flagged modes while computing these averages. Therefore no rejoining and 
smoothing of the data are needed. 
\end{enumerate}
 
We call this algorithm the conditional-counting method. 
It should be pointed out that condition eq.~(23) is applied on each 
scale $j$, and therefore the unwanted modes are flagged on a 
scale-by-scale basis.  If the size of an unwanted data segment is $d$, 
condition eq.~(23) only flags modes $(j,l)$ on scales less than or 
comparable to $d$. We also flag two modes around each unwanted mode 
to reduce any boundary effects of the unwanted chunks.

Since the DWT calculation assumes that the sample is periodized, this may 
cause uncertainty at the boundary of the sample. To reduce this effect, we 
drop five modes neighbouring the boundary of the sample. With this method, 
we can still calculate the power spectrum by the estimators of eq.~(19), 
but the average is not over all modes $l$, but over the un-flagged modes 
only. 

\subsection{Testing the DWT denoising method}

At the first glance, the conditional-counting condition 
eq.~(23) seems to preferentially drop modes in 
the low transmission regions, and may lead to an $f$-dependence of 
power spectrum. To test this problem, we calculate the power spectrum
$P_j$ of Q1700+6419. The conditional-counting parameter $f$ is taken to 
be $f=1,2,3$ and 5. The results of $P_j$ vs $j$ are shown in Fig.~2. 

Fig. 2 shows that the power spectrum $P_j$ is independent of 
the parameter $f$. For other samples, the results are also the same. The 
reason can be seen from eq.~(19), which
shows that the contribution to the power $P_j$ given by mode $(j,l)$ is 
$(\tilde{\epsilon}^F_{jl})^2-(\tilde{\epsilon}^n_{jl})^2$.
The noise subtraction term $(\tilde{\epsilon}^n_{jl})^2$ guarantees 
that the contribution of modes with small $S/N$ to $P_j$ is always small.  
For instance, the modes with negative flux, i.e., the modes with flux having 
the same order of magnitude as noise, the two terms 
$(\tilde{\epsilon}^F_{jl})^2$ and $(\tilde{\epsilon}^n_{jl})^2$ cancel 
each other statistically. Thus, in the range of $f\leq 5$, all the 
flagged modes always have very small or negligible contributions to 
$P_j$. Denoising by thresholding is reliable. The thresholding method checks 
variables mode by mode, and therefore,
can only be effectively applied for a space-scale decomposed field. 
 
\section{Intermittent features of local power}

\subsection{Spikiness of the spatial distribution of local power}

 Eq.~(19) can be rewritten as
\begin{equation}
\label{eq:DWTpjunlocal}
P_j=\frac{1}{N_f}\sum_{l=0}^{N_f}P_{j l},
\end{equation}
and
\begin{equation}
P_{j l}= (\tilde{\epsilon}^F_{j l})^2-(\tilde{\epsilon}^n_{j l})^2,
\end{equation}
where $N_f$ is the number of modes remaining after applying the denoising 
condition of eq.~(23). Eq.~(24) shows that the power spectrum $P_j$ is 
the average of local power $P_{jl}$. For a given $l$, the 
$j$-distribution of $P_{j l}$ is the local power spectrum, i.e., 
the power spectra in the spatial range of $l$. For a given $j$, 
the $l$-distribution of $P_{j l}$ is the spatial distribution of local 
power.

The spatial distribution of local powers is very intuitive for the 
demonstration of intermittency. Fig.~3 shows a typical spatial 
distribution of the local power of the transmitted flux.
The left panels of Fig.~3 are $P_{jl}/P_j$ for $j= 9$ and 10 
($\sim$ 64 km/s and 32 km/s, respectively). The right panels 
represent the corresponding distribution of the phase-randomised 
(PR) data. The PR data are obtained by taking the inverse 
transform of the Fourier coefficients of the original data after 
randomising their phases uniformly over $[0,2\pi]$ without changing 
the amplitudes. This gives rise to a PR field with the same 
unnormalised power spectrum as the original field. The 
mean unnormalised powers of the left and the right panels of 
Fig.~3 are actually the same. Note that, in Fig.~3 large 
spikes corresponding to metal lines have been removed. 

Two main features can be observed in Fig.~3. First, the local power 
distributions of $P_{jl}/P_j$ on $j= 9$ and 10 (left panels) are 
significantly different from their counterparts of the PR 
sample (right panels). The former show spiky structures, while the 
latter are noisy distributions. Therefore, the spikiness arises 
completely from the phase correlation of the Fourier modes. Second, 
the spiky structures are more significant on smaller scales, or
larger $j$. That is, the ratio between the amplitudes of the spikes and
the mean power is higher on smaller scales. 

Although spikes correspond to a large difference in the flux 
$|F(x+\Delta x)- F(x)|$, they are not always related 
to Ly$\alpha$ absorption lines or sharp edges of saturated regions. This 
is because the strength of spikes is measured by the ratio 
$P_{jl}/P_j$. Fig.~2 shows that the mean power of 
transmission flux fluctuations on $j=10$ is about 
$P_{10} \leq 10^{-2}$. Thus, even for a spike as high as 
$P_{jl}/P_j \sim 40$ at $j=10$, the flux difference is only 
$|F(x+L/2^{10}) -F(x)| \simeq 0.6$, which is less than 
$\langle F \rangle=0.7$ and does not require either $F(x+L/2^{10})$ or 
$F(x)$ to  equal zero. Therefore, spikes are different from the Ly$\alpha$ 
absorption lines which correspond to valleys identified by a Voigt 
profile fitting. 

The difference between the spikes and Ly$\alpha$ absorption lines can also
be seen from Fig.~4, which shows the local power $P_{jl}$ on scale 
$j=10$ against the corresponding $\sqrt{2^j/L}\epsilon^F_{jl}$ on 
scales $j=10$ and $j=11$ at the same physical position $l$. Since 
$\sqrt{2^j/L}\epsilon^F_{jl}$ is the flux smoothed on scale $j$, 
absorption lines having a width of the order of  $j$ corresponds to 
$\sqrt{2^j/L}\epsilon^F_{jl} \simeq 0$. Fig.~4 shows the modes with 
top 1 $\%$ power among all modes of the 28 QSOs combined. All these local 
powers $P_{jl}$ are larger than 10$P_j$, and therefore, they 
are spikes. One can see from Fig.~4 that most spikes are not related  
to saturated regions with $F \sim 0$ or $\sqrt{2^j/L}\epsilon_{jl} \sim 0$.

\subsection{PDF of local powers}

By definition, spikes correspond to structures of large  
$|F(x+\Delta x)- F(x)|$. A spiky field
indicates the excess of large transmission fluctuations compared to a 
Gaussian distribution. Thus, the non-Gaussianity of a spiky field
can be described by the PDF or the one-point distribution of local 
power $P_{jl}$. 

To calculate the PDF, we use the 12 redshift ranges of the samples mentioned
in \S 4.1. For each redshift range, we construct an ensemble consisting 
of all $P_{jl}$ from the 28 QSOs, for which the position $l$ is in the 
redshift ranges. Fig.~5 plots the PDFs of $P_{j l}/P_j$ on the scale 
$j=10$ in the 12 redshift ranges. 

If the field $\delta(x)$ is Gaussian, the PDF of 
$\tilde{\epsilon}^F_{j l}$ is also  Gaussian. Thus, the PDF of
$y=P_{j l}/P_j$ should be a $\chi^2(N=1)(y)$ distribution, which is also 
plotted in Fig.~5 (solid line). Comparing to the 
$\chi^2(N=1)$-distribution, the PDFs of the Keck data are generally 
higher at $P_{j l}/P_j \sim 0$,
lower at $P_{j l}/P_j \sim 1$,
and again higher at $P_{j l}/P_j \geq 10$. That is, for most modes the local 
powers are 
low or close to zero, while rare modes have high power (long-tail). 

In Fig.~5, five of the 12 PDFs have tails as long as 
$P_{jl}/P_{j} > 32$, and six have $P_{jl}/P_{j} >10$. 
The number of long-tails cannot directly be used to measure 
the degree of the spikiness, as different PDFs of Fig.~5 are given by 
 different number of independent modes.
Nevertheless, it is clear that all the 
long-tail events have a much larger probability than a Gaussian. 
 
Fig.~6 shows the PDF of the local power $P_{j l}/P_j$ in the 
redshift range $2.8\leq z \leq 3.0$ on scales $j=8$, 9 and 10, 
with the parameter in eq.~(23), f = 1 (top), f = 3 (middle) and 
f = 5 (bottom). We see from Fig. 6 that the observed PDF is independent 
of the parameter $f$. The observed PDF of $P_{j l}/P_j$ at $j=8$ 
is consistent with the $\chi^2$ distributions, but it is
long-tailed for $j=9$ and 10. At $j=10$, the long-tail is given by  
$P_{jl}/P_{j} \simeq 32$. A spike with $P_{j l} > 32 P_j$ 
corresponds to an event $\sim$ 5.6$\sigma$. In Fig.~6, we use 2100 
$P_{jl}/P_{j}$ data points (modes) for the statistics on $j=10$ and 
$f=3$, and therefore the observed probability of the 5.6$\sigma$ event is 
about 4$\times$ 10$^{-4}$. This is much larger than the Gaussian 
probability of an 5.6$\sigma$ event.   

In contrast to Figs.\ 5 and 6, Fig.~7 shows the PDF of $P_{j l}/P_j$  
for a PR sample. As expected the PR field follows 
the $\chi^2(N=1)$-distribution on small scales ($j = 10$). 

Some errors in Figs.\ 5 and 6  might be caused by the 10\% dispersion of the  
mean flux (\S 4.1). As each ensemble in a given redshift range 
contains local power modes $P_{jl}$ from different QSOs, which have 
different mean flux $\langle F(x)\rangle$. However, this error will not
change the intermittent features. This can be seen from Fig. 8, which
gives the PDF of $P_{j l}/P_j $ for {\it one} sample Q0014+8118 at $j=10$. 
Fig.~8 actually shows exactly the same features as in Figs.~5 and 6. 

Moreover, Fig.~8 gives the distribution $P_{j l}/P_j $ for the sample 
Q0014+8118 with and without removing high spikes located in the regions
contaminated by metal lines. The figure shows the existence of a long-tail 
regardless of whether the metal lines and metal line suspects are removed or 
not. Other quasars searched for metal lines show similar results.

Therefore, the long-tail seems to be a permanent feature of the PDF of
local power. It is not caused by noise or other contamination. To measure 
the long-tail we add a lognormal PDF in Fig. 8 given by
\begin{equation}
P(y) =\frac{1}{2^{1/2}\pi^{1/2}y \mu}
 \exp\left \{ -\frac{1}{2}
\left (\frac {\ln y + \mu^2/2}{\mu}\right)^2  \right \}.
\end{equation}
For this PDF, the mean of $y$ is always equal to 1, i.e.,
$\overline{y}=1$. The parameter $\mu$ measures the long 
tail. Larger the $\mu$, longer the tail.
The best fitting to the observed long-tail is $\mu \simeq 1.5$. 
The observed long-tail is longer than lognormal PDFs. The PDF of the flux 
field $\delta(x)$ has a more 
prominent long-tail than a lognormal field.  
  
\section{Intermittency and the precision of power spectrum}

\subsection{Domination of power spectrum by spikes}

As shown by eq.~(24), the power spectra, $P_j$, are given by the mean 
of local powers over independent modes $l=1, ..., N_f$. When the spiky 
features are pronounced, the power of the transmission fluctuations 
is concentrated in the spikes, and therefore, a big fraction of the 
power in 
eq.~(24) actually is dominated by the spikes. 

To demonstrate the spikiness, we calculate the power 
spectrum by averaging local power $P_{j l}$ in the 12 redshift ranges. 
We also calculate the averages over 12 local power ensembles, but dropping 
the top 1\%, 3\% and 5\% of the local power modes. We plot the 
ratio of power after dropping the highest modes to the power without 
dropping any modes. The result is shown in Fig. 9. 

One can see from Fig.~9 that for most cases, dropping the top 5\% modes 
leads to a decrease in $P_j$ by a factor equal to or larger than 2. 
That is, 50\% or even more of the power is given by the top 5\% modes. 
On scale $j=10$, dropping the top 1\% modes leads to a 20\% or more 
decrease in $P_j$. If the field were Gaussian, a top 1\% elimination 
of data would lead to a decrease in $P_j$ of no more than 3\%, and a top
5\% elimination would lead to a decrease of no more than 8\%. 
Therefore, Fig.~9 shows that the power spectrum is substantially 
dependent on the rare events -- high spikes.  The power is concentrated 
in the spikes.
As a consequence, the number of effective modes of the random field 
is significantly reduced, i.e., only the rare modes contribute to the 
measurement of the power spectrum, while other modes are inactive.
 This will leads to uncertainty in the power spectrum. 

\subsection{Uncertainty in the power spectrum of an intermittent field}

Generally, the uncertainty in the power spectrum can be effectively reduced
by increasing the number of independent modes of measurement.
Using the ``fair sample hypothesis'' (Peebles 1980), we first construct
an ensemble of samples by dividing the observed sample into a set of 
$N$ subsamples. We then calculate the mean and variance over the 
ensemble. The precision of the power measurement would then be improved 
if the number $N$ is large, as the error is $\propto \sqrt{1/N}$. If this 
is always true, then we should divide the sample into as many independent 
subsamples as possible.

To detect the power of Ly$\alpha$ transmitted flux fluctuations 
on the scale $k$, the largest possible number $N$ is given by the uncertainty 
relation $\Delta x\Delta k \simeq 2\pi$. Therefore, we should divide the \
1-D flux field $(x_1,x_2)$ into segments with size $\Delta x \geq 2\pi /k $. 
In this case, $\Delta k \leq k$. The segment gives valuable 
information of the power in the band $k \pm \Delta k$. This is just what 
the DWT does. The local power $P_{jl}$, given by DWT, provides the 
largest possible number of modes for detecting power in the band corresponding 
to $k$.   

However the precision improvement factor  $\sqrt{1/N}$ is largely 
based on the central limit theorem. Let us consider eq.~(24) to be 
a definition of the stochastic variables $P_j$ constructed from the stochastic 
variables $P_{jl}$. According to the central limit theorem if all $P_{jl}$ are 
independent variables, having identical PDFs, then $P_j$ will approach a 
Gaussian PDF when $N_f$ is large, regardless of the PDF of $P_{jl}$.
Actually, the convergence of $P_j$ to a Gaussian variable is fast as $N_f$ 
increases. This result ensures that the error of power spectrum 
$P_j$ is basically proportional to the Gaussian factor $\sqrt{1/N_f}$, even 
when the field is nonlinear and non-Gaussian.

Yet, the central limit theorem does not work well with fields having a
divergent ratio between its moments [eqs.(1) or (2)]. Principally, a 
superposition of intermittent fields will also converge to a Gaussian limit.
The error will decrease as $\sqrt{1/N_f}$ as required by the central 
limit theorem. However, this needs a large number $N_f$. This is because  
the central limit theorem relies on the existence of a unique relationship 
between the PDF and moments (Vanmarcke 1983). The PDF of a lognormal field 
is not uniquely determined by its moments (Crow, \& Shimizu 1988). This leads 
to a very slow convergence of the PDF of $P_{j}$ to the limiting Gaussian 
PDF as $N_f$ increases. The PDF of $P_j$ is still long 
tailed if the PDF of $P_{jl}$ is long-tailed, like a lognormal field. The 
number $N_f$ needed for the convergence to a Gaussian $P_j$ can be estimated by 
(Barakat, 1976) as,
\begin{equation}
N_f  \simeq \gamma^2_1,
\end{equation}
where $\gamma_1=(e^{\mu}-1)^{1/2}(e^{\mu}+2)$ is the skewness of the PDF 
eq.~(26). Thus, for $\mu \simeq 1.5$ (\S 5.2), we have 
$N_f \simeq 2.1\times 10^4$. Therefore, the central limit theorem is less 
effective for an intermittent field. For the PDF shown in Fig. 8, the 
precision improvement factor $\sqrt{1/N}$ would not work until $N_f$ is not 
large enough as required by eq.~(27). 

To demonstrate this property, we calculate the mean power and its 1-$\sigma$ 
of Q0014+8118 on the scale $j = 9$. The power is calculated by eq.~(24), and
the 1-$\sigma$ by  
\begin{equation}
\sigma_P= \left [\frac{1}{N_f-1}\sum_{N_f}(P_{jl} -P_j)^2\right]^{1/2}
\end{equation} 
In the redshift range 2.7 - 3.12, there are 397 local power modes $P_{jl}$ 
available. These powers are independent in the sense that the cross 
correlations 
$\langle \tilde{\epsilon}^F_{jl}\tilde{\epsilon}^F_{jl'}\rangle \simeq 0$
if $l\neq l'$ (Pando, Feng, \& Fang 2001). However, the PDF of these  
$P_{jl}$ are close to eq.~(26). The results are given in Table 1. Here
 $N_f$ means that we use only the first $N_f$ local power modes 
$P_{jl}$ out of the total 397 modes. Table 1 shows that 
$\sigma_P$ does not show the $\sqrt{1/N_f}$-dependence. It is 
almost independent of $N_f$. On the other hand, we find the PR 
samples do show a decrease in the error by factor 
$1/\sqrt{N_f}$. 
\begin{table*}
\begin{center}
\centerline{Table 1}
\bigskip
\begin{tabular}{lllll}
\hline\hline
  $N_f$ & $P_j$ & $\sigma_P$ & 95\% confidence & 99\% confidence \\
 \hline
 99     &  0.134 &  0.215 (0.250) &  $<$ 0.844 (0.823) & $<$ 0.999(1.85) \\
 198    &  0.132 &  0.231 (0.246) &  $<$ 0.835 (0.810) & $<$ 1.30 (1.82)\\
 397    &  0.120 &  0.224 (0.223) &  $<$ 0.827 (0.736) & $<$ 1.30 (1.65)\\ 
   \hline
\end{tabular}
\end{center}
\end{table*}

The numbers in the bracket of Table 1 are calculated by using the PDF in 
eq.~(26), 
which gives 
\begin{equation}
\sigma_P= P_j (e^{\mu}-1)^{1/2}, 
\hspace{3mm}
95\% {\rm \ Con.} =  P_j e^{1.96 \mu -\mu^2/2},
\hspace{3mm}
99\% {\rm \  Con.} =  P_j e^{2.58 \mu -\mu^2/2}.
\end{equation} 
We use $\mu=1.5$. All the observed results can roughly be fitted by the 
lognormal PDF eq.~(26). That is, regardless of $N_f$ in the range 99 - 397, 
the PDF of $P_j$ are long-tailed. Again, with the phase randomised 
samples, the PDF of $P_j$ is Gaussian, and with a variance following 
the factor $1/\sqrt{N_f}$.

Therefore, the variance $\sigma_P$ doesn't effectively decrease with 
the increase of the number $N_f$ when $N_f$ is the order of
$ \simeq 10^2$. Table 1 shows also that the values of $\sigma_P$
generally are large. This is directly due to the intermittency, or
the PDF eq.~(26). 

\subsection{Error estimation by subsamples with optimal size}

To reduce the uncertainty or 1-$\sigma$ error of the power spectrum of 
the flux fields, a popular method for estimating the error of power 
spectrum of the flux field is based on subsamples or segments with
optimally selected size. For instance, Croft et al. (2002) used the 
jackknife error estimator, which is found by dividing the sample into $N$ 
subsamples, and computing the 1-$\sigma$ error bars by 
$\sigma_P= [(1/N)\sum_{N}(P_i-\hat{P)}^2]^{1/2}$, where 
$\hat{P}$ is the mean power from the full data samples and $P_i$ is the 
mean power estimated by leaving out the subsample $i$. They used $N=5$ as  
the optimal number. McDonald et al. (2000) used a modified  bootstrap 
error estimator, in which they have re-sampled the data not 
pixel-by-pixel, but segment-by-segment. Each segment has a size of 100 
pixels which is also used as an optimal number. 

The 1-$\sigma$ error given by these methods is indeed much less than the 
$\sigma_P$ listed in Table 1. We need to understand why they yield 
smaller error bars? What is the criterion for selecting the optimal size of subsamples 
and segments? In our view, these problems are not all technical, but rather 
depend on the intermittent nature of the Ly$\alpha$ flux field.  

To demonstrate these problems, we analyse the local power $P_{jl}$ modes. 
The jackknife estimator is equivalent to dividing a total of
$N_f$ local power modes ($P_{jl}$) into $N$ groups (subsamples). Each group 
contains $M$ modes, i.e., $N_f = NM$. The 
$1-\sigma$ uncertainty is calculated by 
\begin{equation}
\sigma_P= \left [\frac{1}{N-1}\sum_{i=1}^{N}(P_{i} -P_j)^2\right]^{1/2},
\end{equation}
and   
\begin{equation}
P_{i}= \frac{1}{M_i}\sum_{l(i)}P^i_{jl},
\end{equation}
where the subscript and superscript $i$ is for the group (subsample) 
$i$, $i=1, \ldots, N$. 
The summation runs over modes $l$ in the group $i$. 
 
Using eqs.~(30) and (31), we again calculate the mean and $\sigma_P$ 
of the $j=9$ power of Q0014+8118. The result is listed in Table~2. Line 1 
of Table 2 is the same as that of Table~1. It is for $N = 397$, and 
$M_i =1$. The lines 2 through 5 are obtained by 
dividing the 397 local power modes ($P_{jl}$) into 40, 20, 10 and 5 
subsamples, 
respectively. When the size of subsamples is 9 and 19 (lines 2 and 3), 
$\sigma_P$ is still large, and the numbers for 95\% Con.\ and 99\% Con.\ 
are also larger than $2\sigma_P$ and $2.5\sigma_P$. Therefore, the PDF 
of $P_i$ is still long-tailed. In the lines 4 and 5, we didn't show
the 95\% Con.\ level, because the total number $N$ of the subsamples is 
only 10 or 5, and is too small to measure the PDF with a resolution $<10 \%$.
The value of 99\% Con.\ indicates that
the PDF of $P_i$ in the lines 4 and 5 probably is no longer long-tailed.
The value of $\sigma_P$ for lines  4 and 5 is much less than lines 1 and 2. 
Thus, with Table 2, the optimal size of the subsamples probably seems to 
be $M \sim 40$ or 80 or $N=10$ or 5. We also calculated powers on other 
scales and found that they have the similar behaviour as the Table 2. 
Therefore, the criterion for the optimal size $M$ might be that the PDF 
of $P_j$ is no longer long-tailed.

From Table 2, we see that the error estimator of subsamples or segments 
with optimal size calculates the dispersion of powers among the 
subsamples and segments only, and therefore, it actually is not a measure 
of the dispersion among all independent modes of the data. If two 
subsamples have the same mean power $P_i$, but very different distributions 
of $P_{jl}$, their contribution to $\sigma_P$ (eq.~(30)) is the same. 
Therefore, $\sigma_P$ does not measure the uncertainty caused by the 
difference between the intermittent subsamples. Only when the two subsamples 
are Gaussian, their statistical properties would be the same if they have 
the same $P_j$ (second moment). Therefore, the error estimator with 
subsamples and segments implicitly {\it assumes} that the subsamples 
are Gaussian. Eq.~(30) with the optimal size subsamples does not give the 
error of the originally intermittent field, but the corresponding Gaussian 
subsamples having the same $P_i$ as the intermittent field. We have 
checked the PDF of $P_{jl}$ of subsamples in lines 4 and 5 of Table 2. There 
are substantially non-Gaussian. 

Therefore, the improvement in the error estimation using optimal size 
subsamples or segments essentially is to replace an intermittent field within 
each subsamples with a Gaussian field. The intermittent features 
of the field are overlooked. Moreover, these errors are parameter 
(optimal size) dependent. According to eq.~(1), the spikiness is scale-dependent, 
and therefore, the optimal size given by Table 2 generally is scale-dependent. 
One parameter option cannot fit the optimal size for all scales. 
\begin{table*}
\begin{center}
\centerline{Table 2}
\bigskip
\begin{tabular}{lllllcl}
\hline\hline
  & $N$  & $M_i$ & $P_j$ & $\sigma_P$ & 95\% Con. & 99\% Con. \\
 \hline
 1 & 397  &     1  &      0.120 & 0.224 &  $<$ 0.827 & $<$ 1.30 \\
 2 & 40   &     9  &      0.123 & 0.073 &  $<$ 0.233 & $<$ 0.282 \\
 3 & 20   &    19  &      0.121 & 0.044 &  $<$ 0.208 & $<$ 0.215 \\  
 4 & 10   &    39  &      0.121 & 0.015 &   - & $<$ 0.143  \\
 5 & 5    &    79  &      0.120 & 0.015 &   - & $<$ 0.143 \\
\hline
\end{tabular}
\end{center}
\end{table*}

\subsection{Error estimation without subsamples}

The analysis of last subsection shows that it is necessary to have
an error estimator for the Ly$\alpha$ flux field, that does
not require any extra parameters like the size of subsamples, and is not 
insensitive to the intermittency of the field. This can be done with
the local power $P_{jl}$. Let us consider eqs.~(24) and (28) once again. 
As mentioned in \S 6.2, the $N_f$ data of $P_{jl}$ can be considered 
independent. They can be used as the parent sample for bootstrap 
resampling. That is, we generate $N_f$ samples having $N_f$ modes by 
drawing $P_{jl}$ from the parent sample with replacement. 
We then calculate the mean and 1-$\sigma$ over $N_f$ realizations.
 
With this method, we again calculate the mean and $\sigma_P$ on  
$j=9$ for Q0014+8118. First, Fig.~10 plots the PDF of the parent 
data of $P_{jl}$ ($j=10$) of Q1103, and two given by bootstrap realizations.
It shows that the PDF of the synthetic data $P^S_{jl}$ of each realization 
is the same as the original $P_{jl}$, i.e., it is long-tailed. 
Therefore, the ensemble of the realizations from the bootstrap resampling
contains all information of the intermittency. 

The results of the mean and $\sigma_P$ are listed in Table 3. Here $N_B$ 
is the number of samples given by bootstrap re-sampling. We see, all the 
results are very stable. Except for $N_f=99$, the relations among  
$\sigma_P$, 95\% Con. and 99\% Con. for $N_f=397$ and 198 are close to 
Gaussian. This is because the total number of modes used in the estimation of  
Table 3 is of the order of $> 10^2\times10^2\simeq 10^4$, which is comparable 
with the number required by the central limit theorem (\S 6.2). The case 
of $N_f=99$ is not large enough, and gives a much larger $\sigma_P$. Therefore, this method illustrates a slow convergence of $P_j$ to the Gaussian limit 
due to intermittency.
\begin{table*}
\begin{center}
\centerline{Table 3}
\bigskip
\begin{tabular}{llllll}
\hline\hline
  $N_f$ & $N_B$  & $P_j$ & $\sigma_P$ & 95\% Con. & 99\% Con. \\
 \hline
  397  &  397    & 0.120 & 0.0110  & $<$ 0.140 & $<$  0.153 \\
  198  &  198    & 0.119 & 0.0159  & $<$ 0.145 & $<$  0.162 \\
  99   &   99    & 0.120 & 0.0231  & $<$ 0.167 & $<$  0.185 \\
\hline
\end{tabular}
\end{center}
\end{table*}

As a final result, we show in Fig.~11 the mean power $P_j$ and their 
error bars on scales $j=8$, 9 and 10 given by the estimator of the 
bootstrap resampling developed above. The powers in Fig.~11 are shown in each 
redshift bin of the Keck data.  As the field is highly non-Gaussian,
we use confidence level to describe the uncertainty range. The error bars 
are the 99\% confidence range of the bootstrap resampling. 
The error bars are independent of the parameter $f$, which is taken to 
be 1 (circle), 3 (triangle) and 5 (pentagon). 
 
Fig.~11 shows clearly the effect of intermittency. The error bars at 
 $z\simeq 1.7$ and 1.9 are much larger than others, because these
samples are not large enough. For redshift bins, for which the number of 
independent modes matches the condition of the CLT convergence, this 
bootstrap resampling method effectively reduces the uncertainty 
in the power spectrum without introducing extra parameters.
    
\section{Discussion and conclusions}

The most popular statistical measure in large scale structure study 
is the power spectrum. For Gaussian fields, there are many effective and 
successful algorithms of calculating the power spectrum and estimating their
errors. The power spectrum in the nonlinear regime is also important not 
only for constraining cosmological parameters, but determining the initial 
conditions for the simulations of galaxy formation. However, the effects 
of non-Gaussianity upon the power spectrum have not been fully studied
yet.  

We studied the effect of the non-Gaussianity of the field of QSO Ly$\alpha$ 
transmitted flux fluctuations on the estimation of their power spectrum 
and errors. The flux field is intermittent. Generally speaking, intermittency 
poses problems in the detection of power spectrum, as a large fraction of 
the power of the transmission fluctuations is concentrated in rare and 
improbable events: high spikes. Thus, the power spectrum is sensitive to 
small probability events. Therefore, the dispersion among the different 
spatial regions is large. This property directly challenges the 
application of the ``fair sample hypothesis'' (Peebles 1980) which assumes 
that a part of the universe is a fair sample of the whole and ensemble 
averages can be calculated by spatial averages. That is, the spatial average
will not converge to the ensemble averages if the dispersion among the 
different spatial regions is large. Mathematically, this is 
shown by the slow convergence of $P_j$ to the Gaussian limit required by the 
central
limit theorem, if the PDF of the field is long-tailed due to intermittency.
 
To reasonably estimate the uncertainty of the power spectrum, 
we should carefully analyse the convergence of the 
data set considered, to the CLT. With this result, we show that some 
conventional methods 
essentially estimate the errors by the dispersion of the powers among 
subsamples, ignoring the dispersion among all the independent modes of 
the data. They do not measure the error of the non-Gaussian or intermittent 
field, but a Gaussian field. The error given by this method is 
parameter-dependent.

With analysis of the CLT convergence, we proposed an error estimator for the 
power spectrum. It is based on a 
bootstrap resampling among the local powers of all the  modes. This estimation 
doesn't need any extra parameter like the size of subsamples, 
and doesn't ignore the intermittency of the fields. The powers and 
errors for 28 transmitted flux samples are calculated with this method.  
This result shows the effect of intermittency on power spectrum, and 
gives a more effective estimation of errors than the ``optimal" size 
method, especially when the number of independent modes matches the condition 
of the CLT convergence.   

We studied in this paper only the unnormalised power spectrum eq.~(18), but not
the normalised power spectrum eq.~(17). If the normalisation is given by a 
constant $\langle{F(\lambda)}\rangle^2$, all the results for unnormalised 
power spectrum hold for the normalised power 
spectrum. If the background 
$\langle{F(\lambda)}\rangle^2$ is not constant but position dependent, the 
local fluctuations will couple with the background when the field is 
non-Gaussian (Jamkhedkar, Bi, \& Fang, 2001). This case should be studied
in detail for the detection and error estimation of the normalised power spectrum. 

\section*{Acknowledgements}

We thank J.~Pando and H.~G.~Bi for their comments. LLF acknowledges 
support from the National Science Foundation of China (NSFC) and 
National Key Basic Research Science Foundation. PJ was supported by the 
Deans Dissertation Fellowship from the University of Arizona.
PJ would also like to thank Drs.~Stein and Shupe and the Department of 
Physics at the University of Arizona for their support.

\newpage

\begin{figure}
\centering
\includegraphics[width=6in]{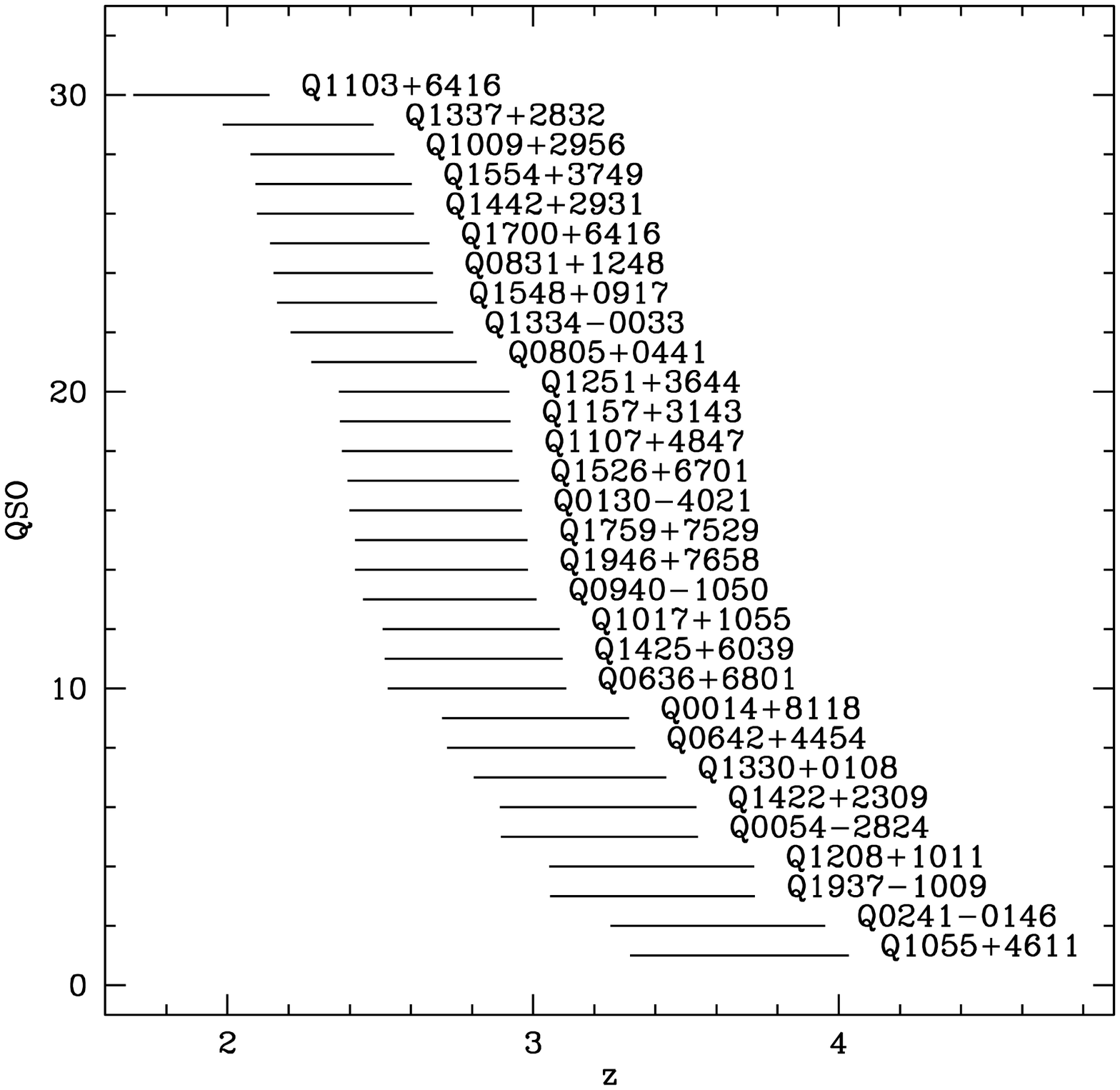}
\caption{Redshift range of the transmission flux of the 30 Keck QSO 
  absorption spectra.}
\end{figure}

\newpage

\begin{figure}
\centering
\includegraphics[width=6in]{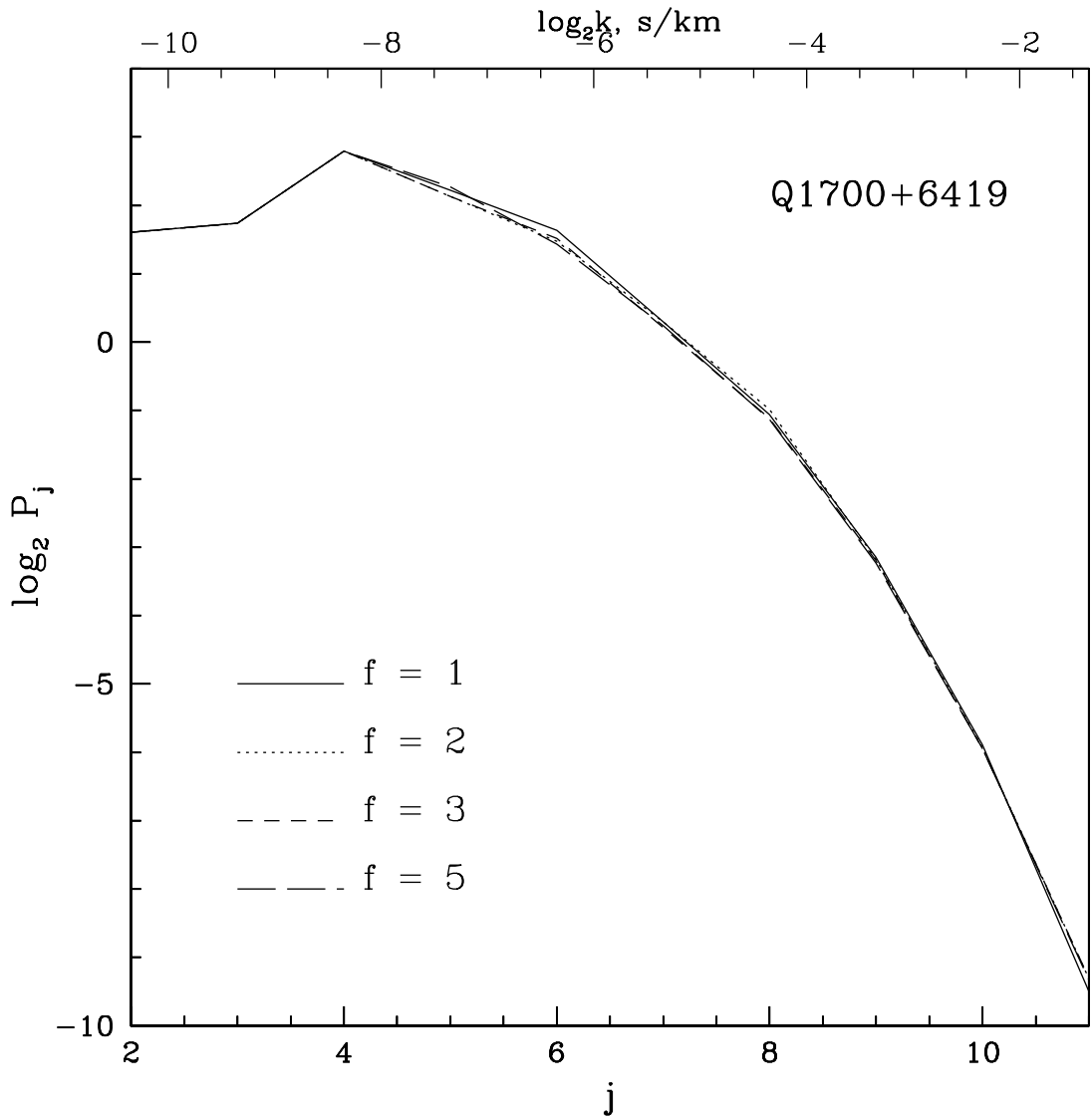}
\caption{The DWT power spectra $P_j$ of Q1700+6419. The parameter $f$ 
for the conditional counting is taken to be 1, 2, 3 and 5. From \S 4.1, 
the scale $j$ corresponds to 
$\Delta v=2c\{1-\exp[-(1/2)2^{13-j}\delta v/c] \}\simeq 2^{13-j}\delta v$,
and $k =2\pi/ 2^{13-j}\delta v$~s/km.} 
\end{figure}


\begin{figure}
\centering
\includegraphics[width=6in]{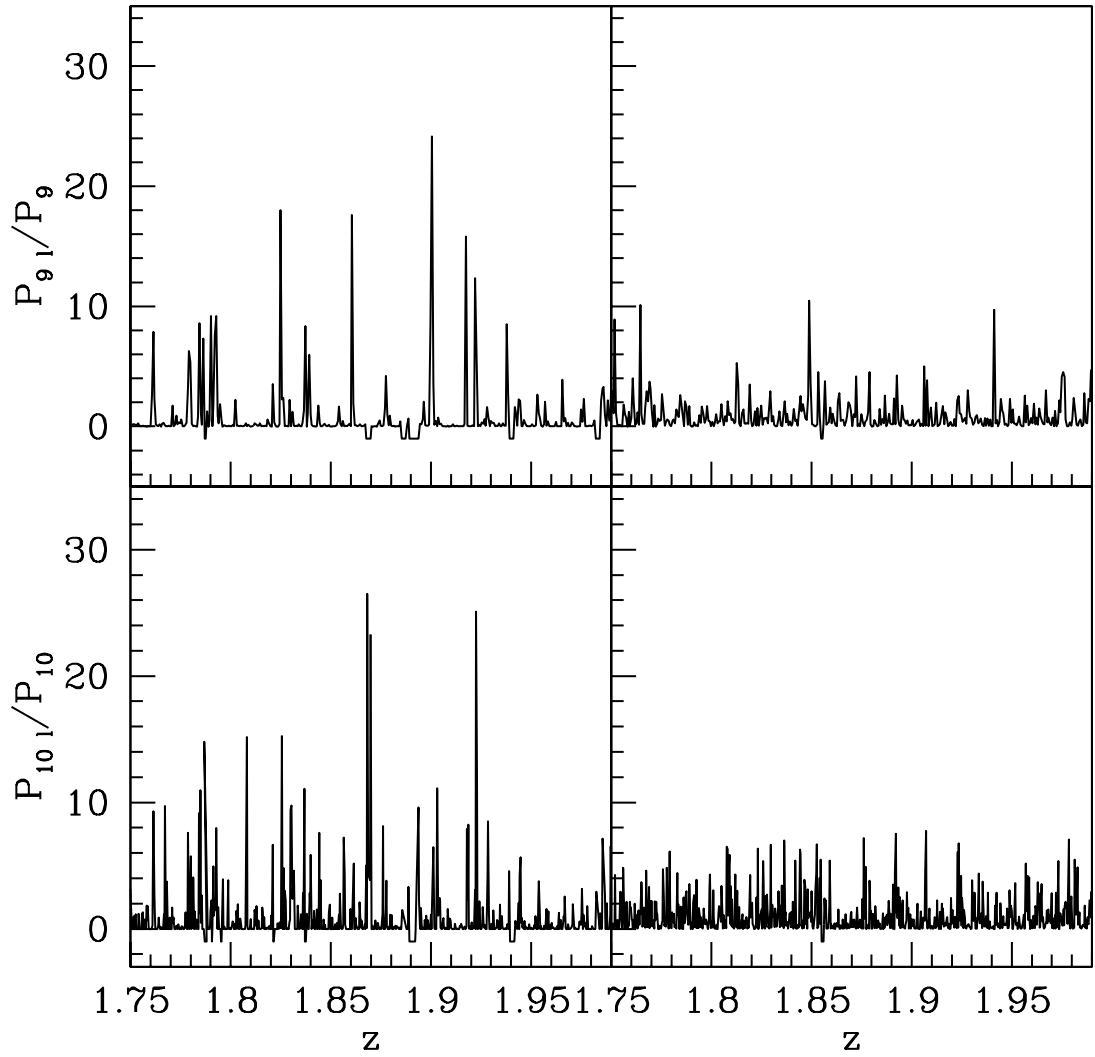}
\caption{A section of 
the spatial distribution of local powers, $P_{j l}/P_j$ on scales 
$j=9$ and 10 (corresponding to $\sim 64$ and 32 km/s) for the sample 
Q1103+6416. The real field is shown in the left panels and the right panels
represent the PR field. Large spikes associated with metal lines have been 
removed.}
\end{figure}

\begin{figure}
\centering
\includegraphics[width=6in]{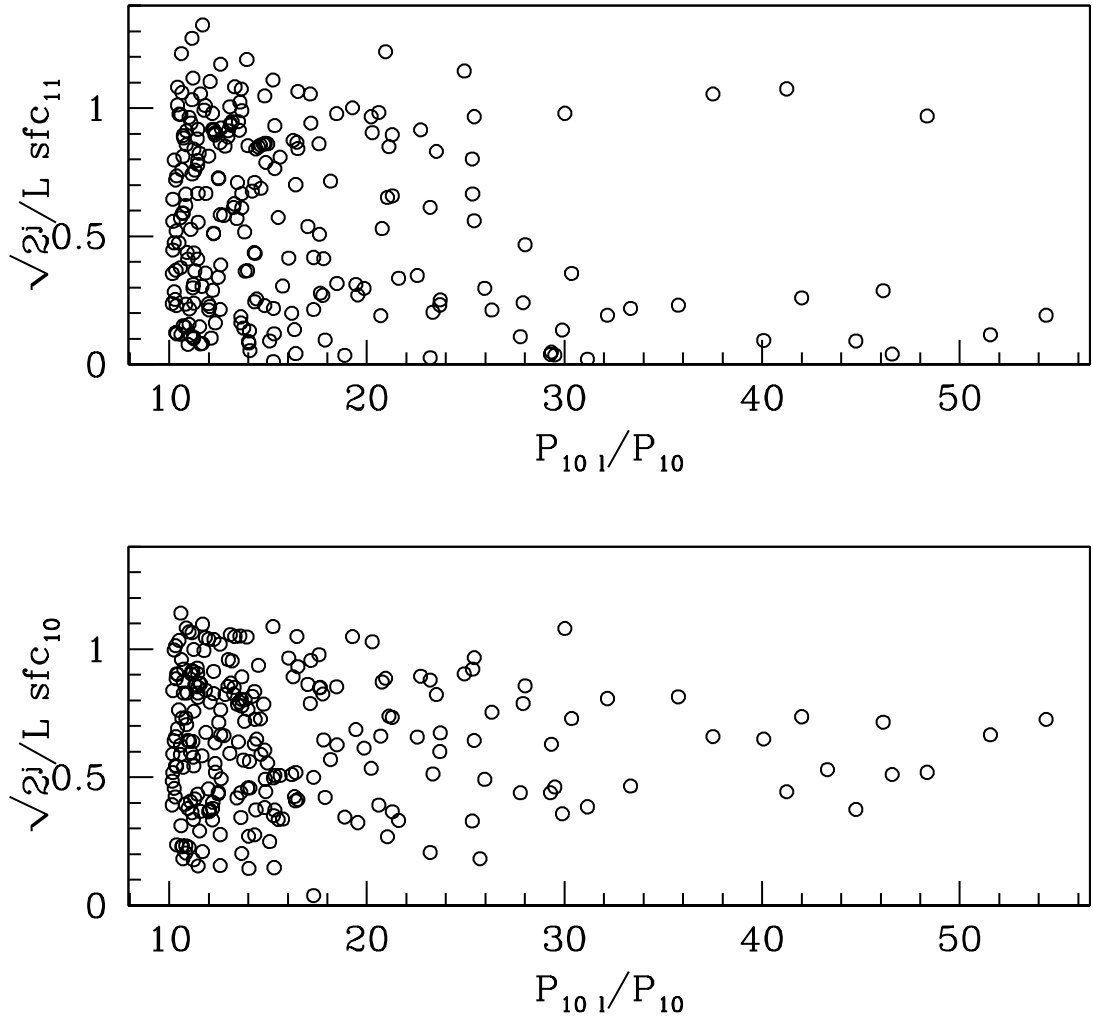}
\caption
{The top 1\% local powers on scale $j = 10$ and their 
corresponding smoothed fluxes (SFCs) on scales $j =11$ (top panel) 
and $10$ (bottom panel).
}  
\end{figure}

\begin{figure}
\centering
\includegraphics[width=6in]{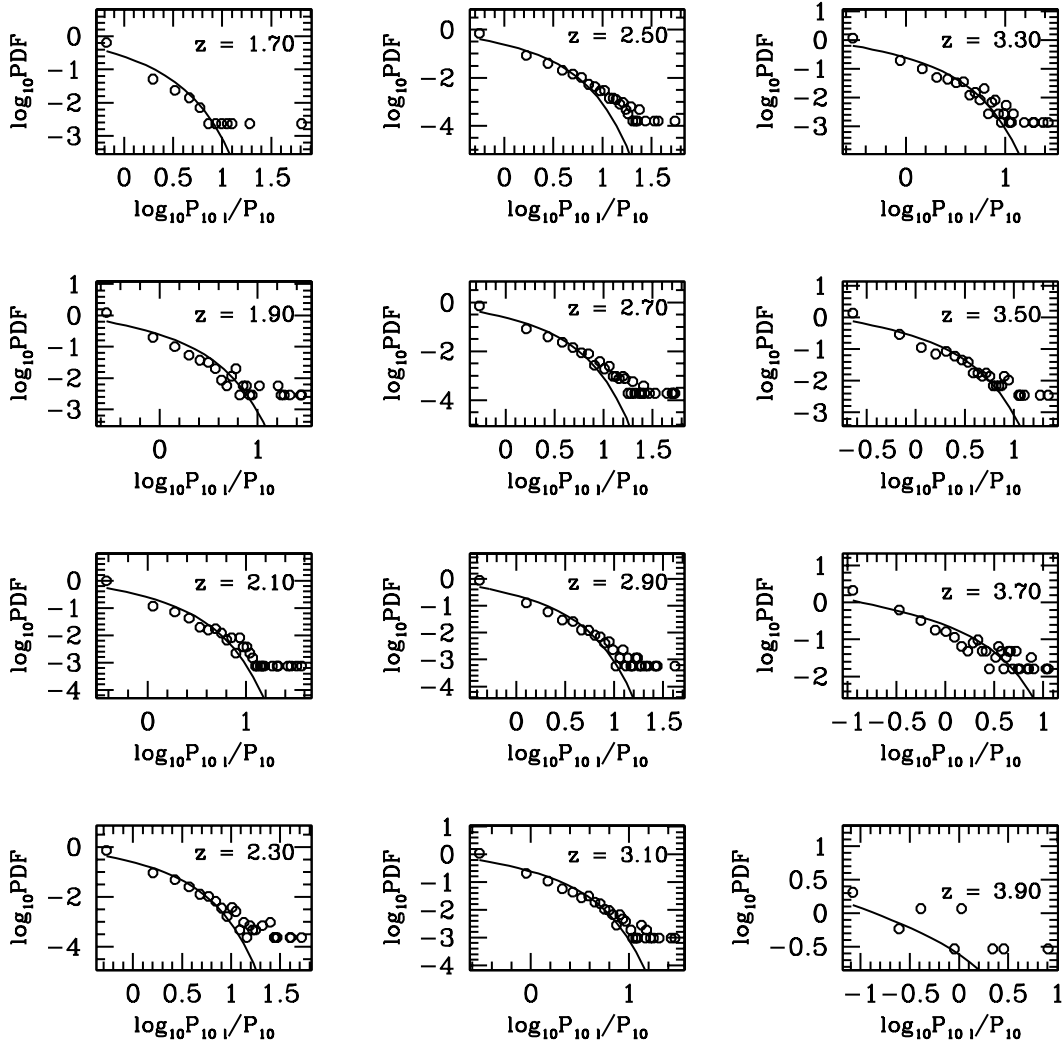}
\caption{The PDFs of $P_{jl}/P_j$ of real samples on the scale $j=10$
  in 12  redshift ranges $z=1.6+n\times 0.2$ to $1.6 + 
  (n+1)\times 0.2 $, $n=0$,\ldots,11. The solid lines are the
  $\chi^2(N=1)$ distribution.}
\end{figure}

\begin{figure}
\centering
\includegraphics[width=6in]{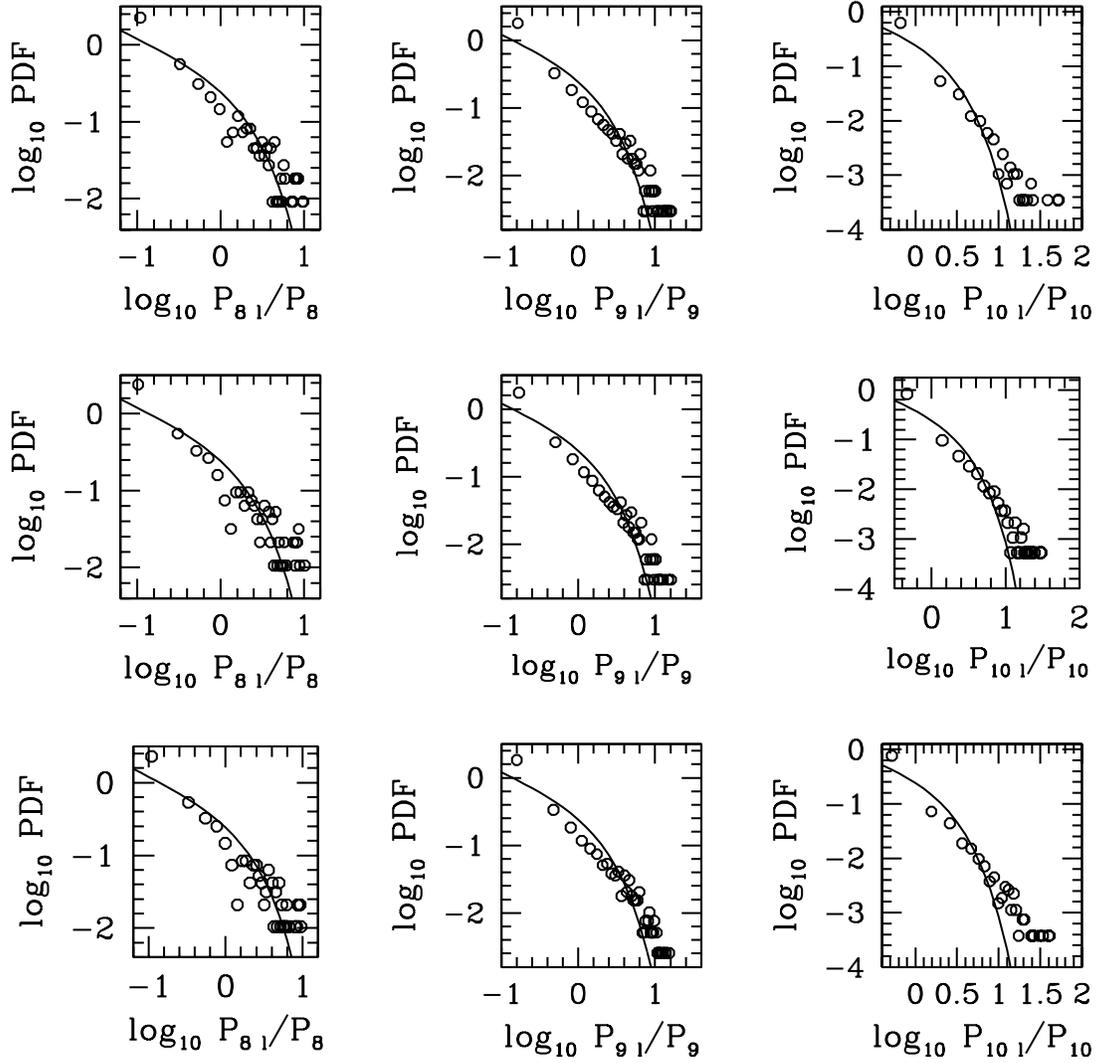}
\caption
{The PDF of $P_{jl}/P_j$ of real samples on scales 
  $j=8,$ 9 and 10 in redshift range $z=2.8$ to 3.0. The parameter $f$ 
  is taken to be 1 (top), 3 (middle) and 5 (bottom). The solid lines are the
  $\chi^2(N=1)$ distribution.}
\end{figure}


\begin{figure}
\centering
\includegraphics[width=6in]{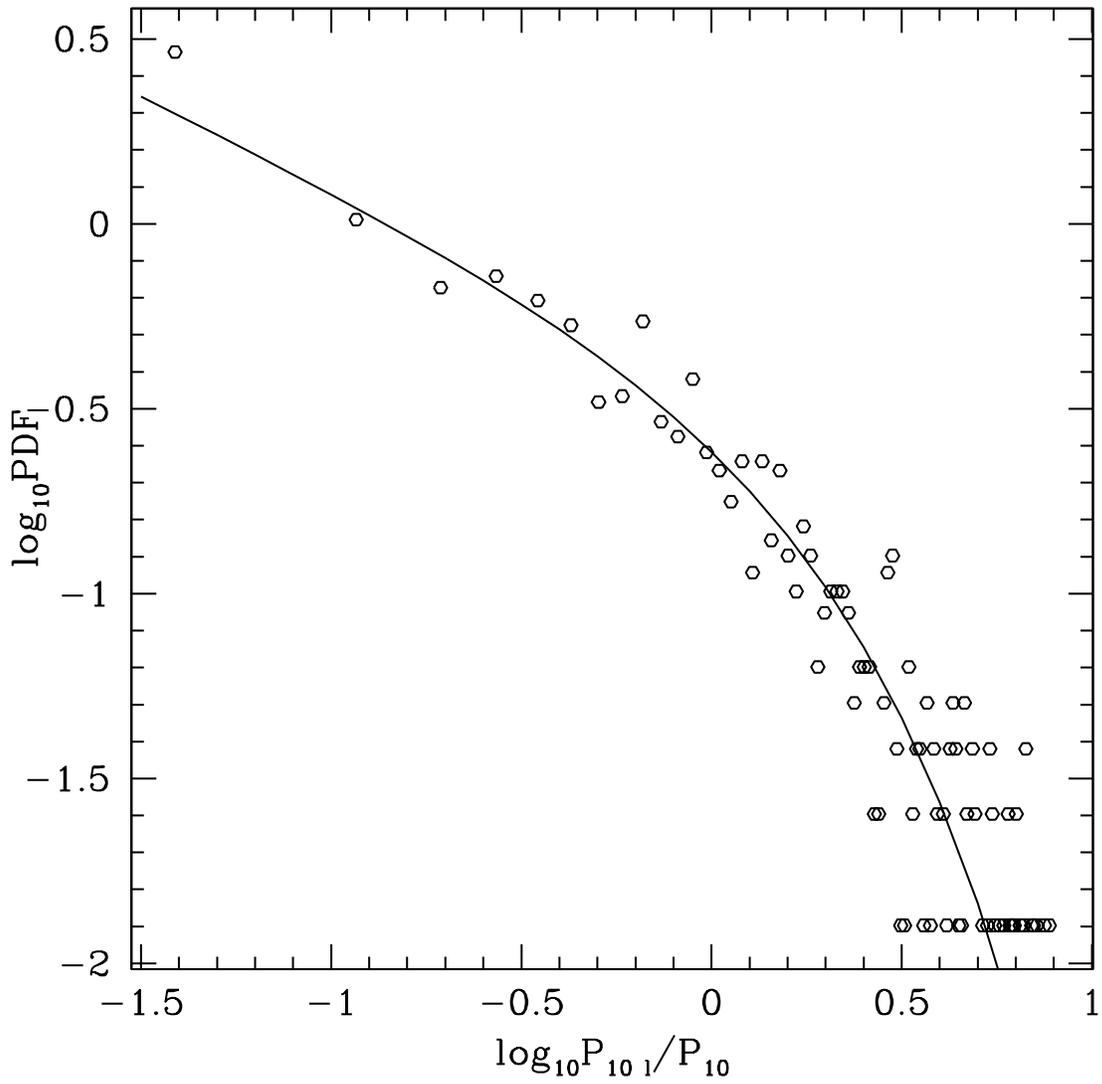}
\caption{The PDF of $P_{jl}/P_j$ of the PR sample of Q1103+6416 on 
  scales $j=10$. The solid line is the
  $\chi^2(N=1)$ distribution.}
\end{figure}


\begin{figure}
\centering
\includegraphics[width=6in]{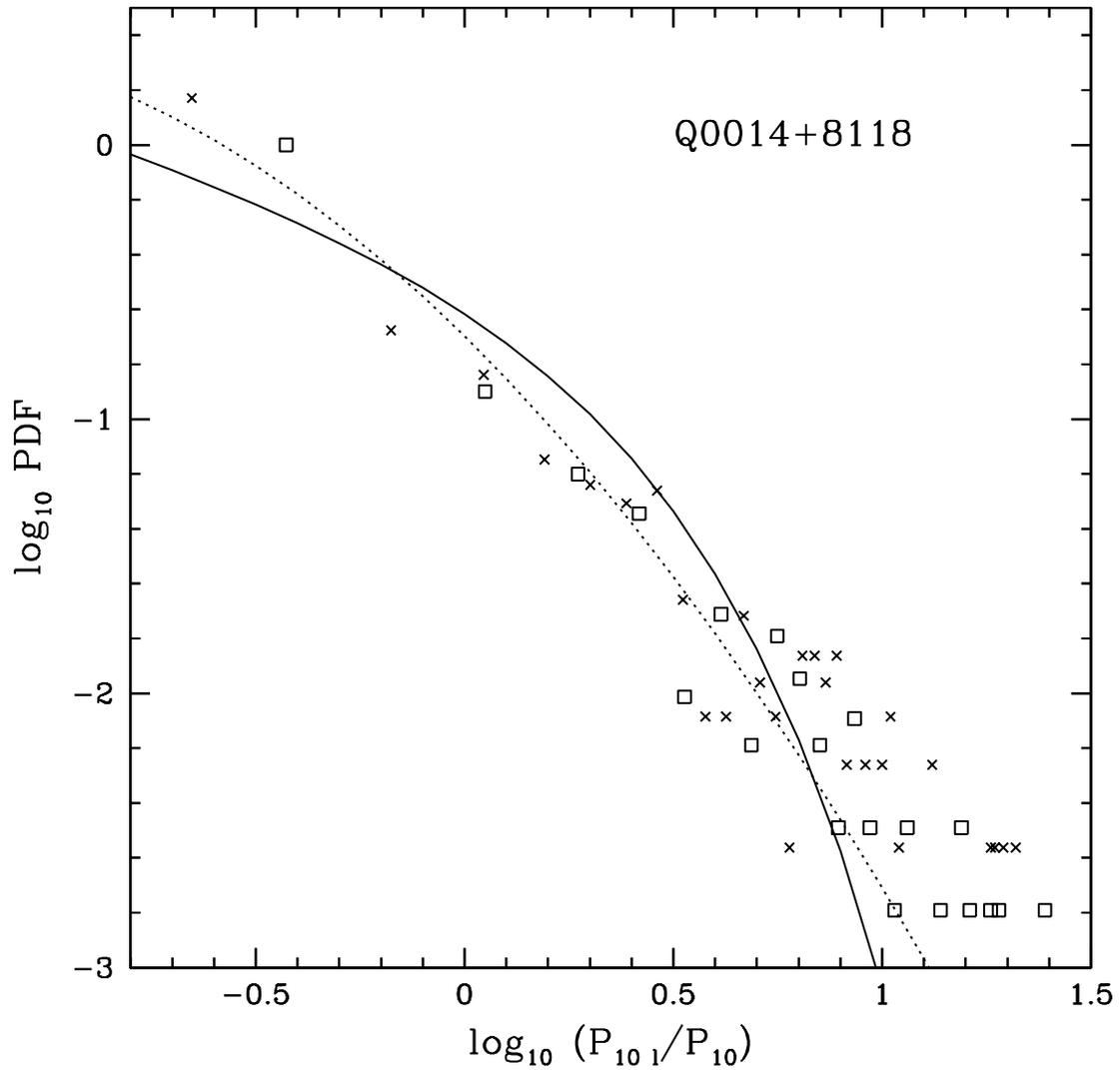}
\caption{The PDF of $P_{j l}/P_j$ for Q0014+8118 and $f = 3$
without the removal of metal lines (squares), and with the removal of metal 
lines and 
metal line suspects from the big spikes (crosses). The solid line is the
  $\chi^2(N=1)$ distribution. The dotted line is the lognormal distribution
(eq.~(26)) with $\mu = 1.5$.
}
\end{figure}

\begin{figure}
\centering
\includegraphics[width=6in]{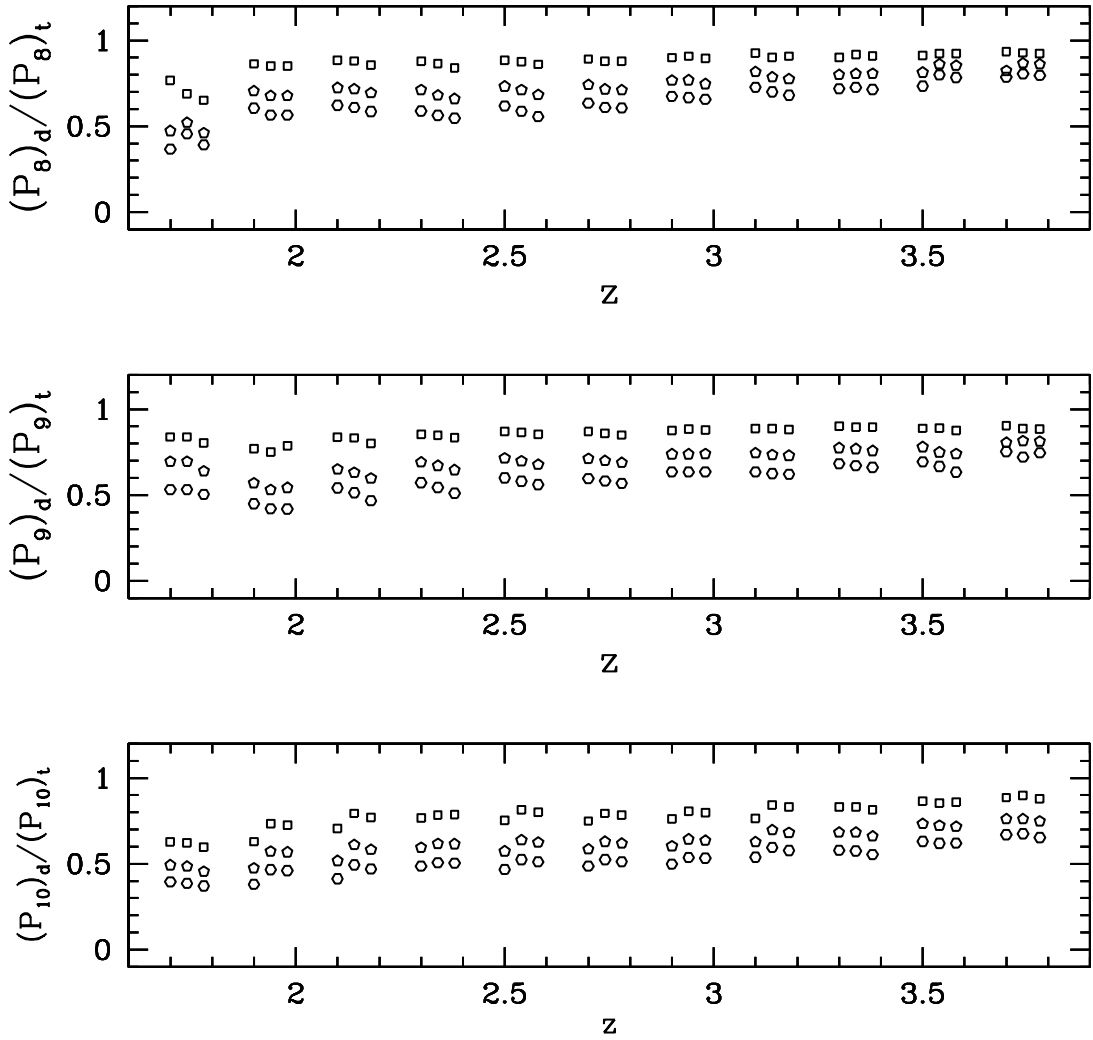}
\caption{Ratio of power $(P_{jl})_d$ with top modes dropped 
to the total power vs. redshift of real data on scales 
  $j=8,$ 9 and 10. Square, pentagon and hexagon are, respectively, the ratio
of powers when the top 1\%, 3\% and 5\% data are dropped.}
\end{figure}

\begin{figure}
\centering
\includegraphics[width=6in]{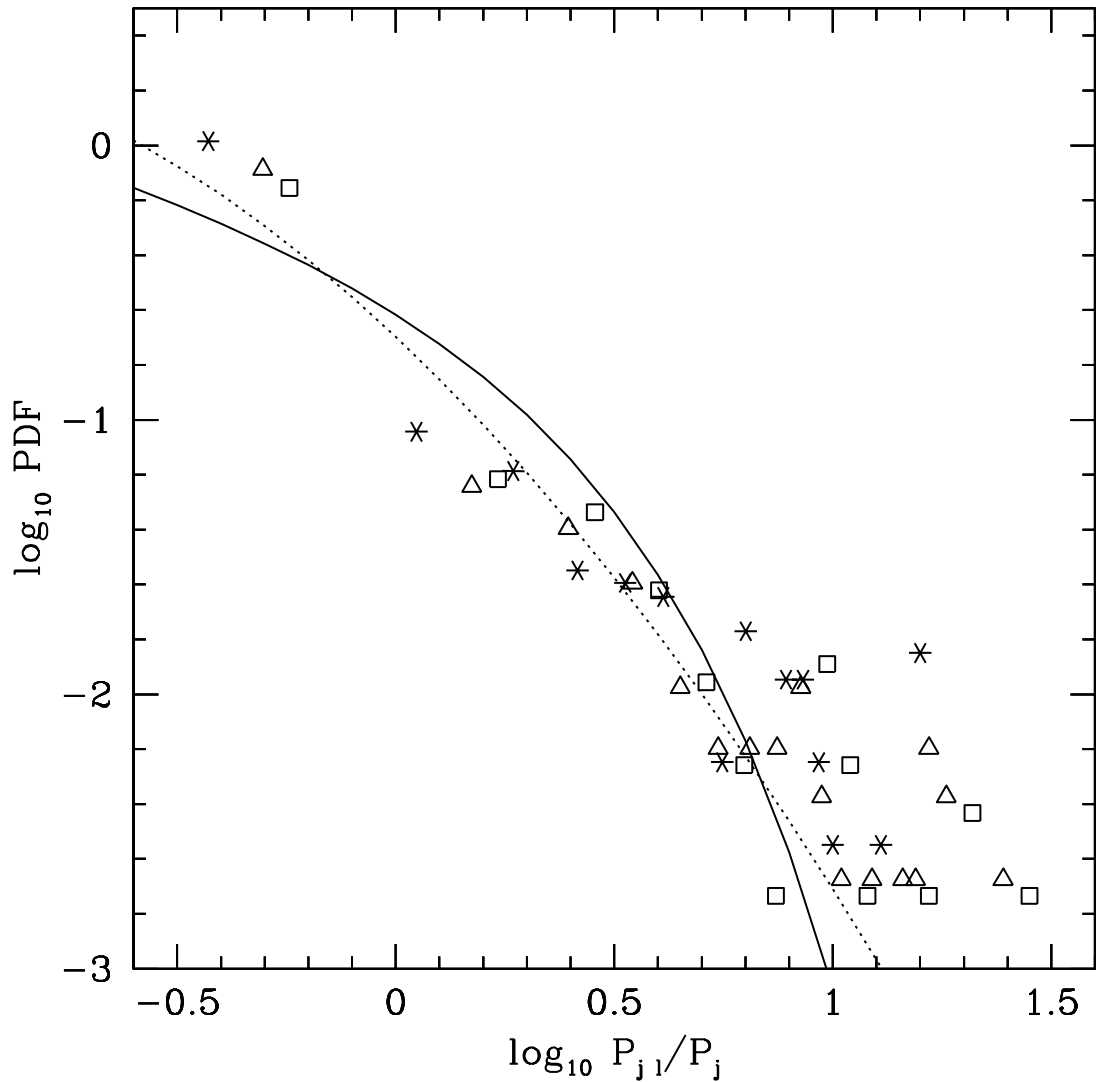}
\caption{The PDF of $P_{jl}/P_{j}$ for Q1103 on scale $j=9$. Triangles
are for real data. Squares and crosses correspond to two bootstrap
realizations. The solid line is the
  $\chi^2(N=1)$ distribution. The dotted line is the lognormal distribution
(eq.~(26)) with $\mu = 1.5$.
}
\end{figure}

\begin{figure}
\centering
\includegraphics[width=6in]{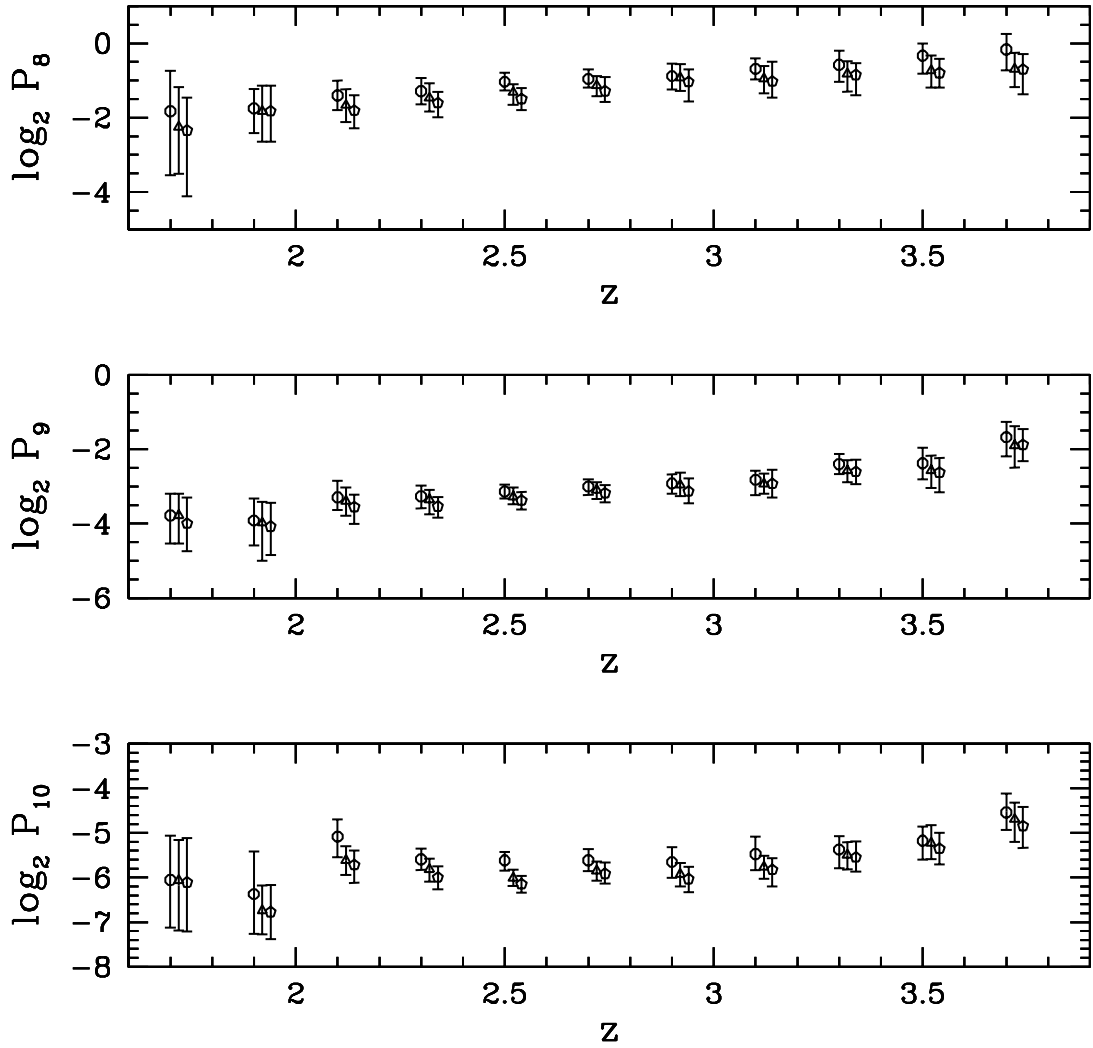}
\caption{The power spectrum, $P_j$ on scales $j=8$, 9 and 10, in each 
  redshift bin of the Keck data. Error bars are the 99\% confidence given by
  the bootstrap resampling. The parameter $f$ is taken to be 1 (circle), 3 
   (triangle) and 5 (pentagon).}
\end{figure}

\end{document}